\documentclass[12pt]{amsart}
\usepackage{amscd}
\usepackage{amsfonts}
\usepackage{epsf}
\usepackage{epic}
\usepackage{eepic}
\usepackage{latexsym}

\begin{document}

\title[On Anyonic Propagators]
{On Anyonic Propagators}
\author{Wellington da Cruz}
\address{Departamento de Fisica, Universidade Estadual de Londrina,
Caixa Postal 6001, Londrina-Parana, Brazil\,\, {\em E-mail address:}
{\rm wdacruz@exatas.uel.br}}

\date{December, 1999}

\maketitle

\begin{abstract}

We consider a simple action for a fractional spin particle 
and a path integral representation for the
propagator is obtained in a gauge such that the 
constraint embodied in the Lagrangian is not an obstacle.
We obtain a 
propagator for the particle in a constant electromagnetic
field via the
path integral representation over velocities, which is
characterized 
by arbitrary boundary conditions and the absence of time 
 derivatives following integration over bosonic variables.

\end{abstract}
\vspace{0.3cm}

There are various actions in the literature\cite{R1}
for particles 
with fractional spins, called anyons\cite{R2}, that can be 
present in nature and explain phenomena such as the
 fractional quantum
 Hall effect and high temperature 
superconductivity\cite{R3}. Such models have been
 constructed in at least 
two ways: one is coupling to some matter field a statistical 
 field of the 
Chern-Simons type that changes the statistics; and the other 
is the group-theoretical 
method that implies a classical mechanical model
 with a Poincar\'e 
invariant 
Lagrangian which is quantized\cite{R4,R5}. In this Letter, we 
consider a simple action 
obtained by the latter approach with minimal extension of 
the phase space that preserves all canonical structure of the 
space-time and spin algebra.
 We propose a path integral representation for the propagator 
in a  convenient gauge. In\cite{R4}, the 
following continuous family of 
Lagrangians for free anyons was obtained

$$
L_{s}=\frac{1}{2e}\left({\dot {x}_{\mu}}
-\frac{s\;{\dot q}\;
f^{\prime}_{\mu}}
{m\;f.\;{\dot x}}\right)^{2}+\frac{e\;m^2}{2}
-\frac{\alpha\; m\; e\;
 {\dot q}}{f.\;{\dot x}},
\eqno{(1)}
$$
where $ m $ and $\alpha$ are 
respectively the mass and
the helicity 
of the anyon, and $ s$ is an arbitrary constant
 deduced from the 
Casimir of the spin algebra, $ S^2=-s^2$.
 The functions $f_{\mu}$ satisfy the non-linear
  differential equation
 
$$
f_{\mu}=\epsilon_{\mu\alpha\beta}f^{\prime\alpha}f^{\beta}
\eqno{(2)}
$$
and we obtain the relations
 
$$
 f^2=0,\;f^{\prime 2}=-1,\;f^{\prime\prime}.\;f=1,
\eqno{(3)}
$$
 with $ f^{\prime}=\frac{df}{dq}$. On the
  other hand,
  we have that 
 $ S_{\mu}=\pi\;f_{\mu}+s\;f_{\mu}^{\prime}$, so we get,
  $ S_{\mu}\;
 f^{\prime\mu}=-s$,$\;$ $ S_{\mu}\;f^{\mu}=0$,$\;$
 $ S_{\mu}^{\prime}\;f^{\mu}=s$,$\;$
 $ S_{\mu}^{\prime}\;f^{\prime\mu}=-\pi$. The equation $
  {\dot f}-{\dot q}f^{\prime}=0$ is equivalent to Eq. (2)
   and we have also

$$ 
 {\dot f}^{\prime}
 -\frac{\partial{\dot q}}{\partial{q}} f^{\prime}
 -{\dot q}\frac{\partial{f^{\prime}}}{\partial{q}}=0 
$$ 
and $f_{\mu}=a_{\mu}q^2+b_{\mu}q+c_{\mu}$, 
 with $a_{\mu}=(1,0,-1)
 $,$\;$$b_{\mu}=(-1,1,1)$,$\;$$c_{\mu}=\frac{1}{2}(1,-1,0)$.
  In this way,
  we extract for $q$ not constant the condition
   $\frac{\partial{\dot q}}
  {\partial{q}}=0$. The Lagrangian 
 is subject to the constraint, $ \sigma\;(f.\;{\dot x})=e\;{\dot q}$,
  and with 
 ${\dot x}_{\mu}=e\;p_{\mu}+\sigma \;S_{\mu}$, we obtain 
 $ {\dot q}=\sigma \; (f\;.\;p)$ 
 which satisfies and gives us that $ f^{\prime}\;.\;p=0$. Then, 
 $ {\dot x}\;.\;f^{\prime}=-\sigma \;s$ and 
 $ \pi \;(f\;.\;p)=-\alpha \; m$,
 that is, $ S_{\mu}\;p^{\mu}+\alpha m=0$ and $ {\dot x}\;.\;f
 =e\;f^{\mu}\;p_{\mu}$, and in the gauge
  ${\dot {\sigma}}=0,\;\
 {\dot {e}}=0$, we have ${\dot q}=m\;f.\;{\dot x}$,
  so $ \sigma=e\;m$
  and  $ p^2-m^2=0 $.  
 
 Now, we consider the path integral 
 representation for 
 the anyonic propagator with the action, ${\mathcal A}=
 \int_{0}^{1} L_{s}\;d\tau$.
 In that gauge, we verify 
 that the continuous family of Lagrangians appears as
  
$$
L_{s}=\frac{{\dot x}^2}{2e}+\frac{s^2}{2e}
  +\frac{e}{2}(m^2-2m^2\alpha).
\eqno{(4)}
$$
  
In this way, we propose the path integral 
  representation for the free fractional spin particle as
 
$$
{\mathcal F}(x_{out},x_{in})=\frac{\imath}{2}\int_{0}^{\infty} 
de_{0}\int_{x_{in}}^{x_{out}}
{\mathcal D}x\;{\mathcal D}e\; M(e)\;\delta({\dot e})
$$
$$
\times\exp\left\{-\frac{i}{2}\int_{0}^{1}
\left[\frac{{\dot x}^2}{e}+
e{\mathcal M}^2+\frac{s^2}{e}\right]
d\tau\right\},
\eqno{(5)}
$$
where $ {\mathcal M}^2=m^2-2m^2\alpha$ and the delta
 functional takes into account the gauge, ${\dot{e}}=0$. 
We define the measure

$$ 
M(e)=\int {\mathcal D}p\;\exp\left\{\frac{\imath}{2}
\int_{0}^{1}e\;p^2\; 
d\tau\right\},
\eqno{(6)}
$$
such that, making the sustitution

$$
{\sqrt {e}}p\rightarrow p, \; 
\frac{x-x_{in}-\tau{\Delta x}}{\sqrt e} \rightarrow x,\;
{\Delta x}=x_{out}-x_{in},
$$
we obtain the new boundary conditions
 $x(0)=0=x(1)$. So, 
in Eq. (5), we have the factor:

$$
I=\frac{\imath}{2}\int_{0}^{0} {\mathcal D}x\; {\mathcal D}p\; 
\exp\left\{\frac{\imath}
{2}\int_{0}^{1}(p^2-{\dot x}^2)d\tau\right\}
$$
$$
=\sqrt{\frac{\imath}{4(2\pi)^3}}.
\eqno{(7)}
$$
Given that

$$
\int {\mathcal D}e\; \delta\left({\dot e}\right){\mathcal G}(e)
={\mathcal G}(e_{0}),
$$
we arrive at the path integral representation for 
the propagator

$$
{\mathcal F}(x_{out},x_{in})=\sqrt{\frac{\imath}{4(2\pi)^3}}
\int_{0}^{\infty}
\frac{de_{0}}{\sqrt{e_{0}^3}}\exp\left\{-\frac{\imath}{2}
\left[e_{0}
{\mathcal M}^2+\frac{{\Delta x}^2}{e_{0}}-\frac{s^2}
{e_{0}}\right]
\right\}
\eqno{(8)}
$$
and $s^2>0$ for the causal propagator. This 
expression 
was obtained along the same lines as those used elsewhere 
for the scalar particle\cite{R6}. On 
the other hand, in momentum space, the anyonic 
propagator has the form\cite{R7}

$$
{\tilde{\mathcal F}}(p)=\frac{1}{(p_{\mu}-mS_{\mu})^2-m^2}.
\eqno{(9)}
$$

Now, we consider the light-like case, for the anyon in a constant
 electromagnetic field $ A_{\mu}=-\frac{1}{2}F_{\mu\nu}x^{\nu}$. 
 Following\cite{R8}, we perform the calculation in the representation 
over velocities, which is characterized by the absence
of derivatives in time after integration over bosonic variables and 
arbitrary boundary conditions. The propagator, in a first step, can 
be written as

$$
{\mathcal F}(x_{out},x_{in})=\sqrt{\frac{\imath}{4(2\pi)^3}}
\int_{0}^{\infty}\frac{de_{0}}{{\sqrt{e_{0}^3}}}
\exp\left[-\frac{\imath}{2}\left(e_{0}{\mathcal M}^2+
\frac{{\Delta x}^2}{e_{0}}-\frac{s^2}{e_{0}}\right)\right]
$$
$$
\times\int {\mathcal D}v\; \delta^3\left(\int v d\tau\right)
\exp\left\{\imath\int\left[-\frac{v^2}{2}
-g({\sqrt e_{0}}v+{\Delta x})A\right.\right.
$$
$$
\left.\left.
\times\left({\sqrt e_{0}}\int_{0}^{\tau} v(\tau^{\prime})
d\tau^{\prime}+x_{in}
+\tau{\Delta x}\right)\right]d\tau
\right\},
\eqno{(10)}
$$
where the new measure ${\mathcal D}v$ has the form

$$
{\mathcal D}v=Dv\left[\int Dv\; \delta^3\left(\int v
d\tau\right)
\exp\left\{-\imath\int \frac{v^2}{2}\;d\tau\right\}\right]^{-1}.
\eqno{(11)}
$$

For the particle in a constant electromagnetic field we 
can obtain for the propagator 

$$
{\mathcal F}(x_{out},x_{in})=\sqrt{\frac{\imath}{4(2\pi)^3}}
\int_{0}^{\infty}
\frac{de_{0}}{\sqrt{e_{0}^3}}
$$
$$
\times\left[\frac{{\rm det}\; L(g)}{{\rm det}\; L(0)}
\right]^{-\frac{1}{2}}
\left[\frac{{\rm det}\; Q(g)}{{\rm det }\; Q(0)}
\right]^{-\frac{1}{2}}
$$
$$
\times\exp\left\{\frac{\imath}{2}gx_{out}Fx_{in}
-\frac{\imath}{2}e_{0}{\mathcal M}^2-\frac{\imath}{2}\frac{s^2}{e_{0}}
-\frac{\imath}{2e_{0}}{\Delta x}
Q^{-1}{\Delta x}\right\},
\eqno{(12)}
$$
where 
 
$$
L_{\mu\nu}(\tau,\tau^{\prime})=\eta_{\mu\nu}\delta
(\tau-\tau^{\prime})
-\frac{1}{2}ge_{0}F_{\mu\nu}\epsilon(\tau-\tau^{\prime}),
$$
$$
Q=\frac{2}
{ge_{0}F}\tanh\frac{ge_{0}F}{2}.
$$

Thus we have, finally, the expression for the propagator as 

$$
{\mathcal F}(x_{out},x_{in})=\sqrt{\frac{\imath}{4(2\pi)^3}}
\int_{0}^{\infty}\frac{de_{0}}{\sqrt{e_{0}^3}}
\left[{\rm det}\frac{\sinh\frac{ge_{0}F}{2}}
{\frac{ge_{0}F}{2}}\right]^{-\frac{1}{2}}
$$
$$
\times\exp\left\{\frac{\imath}{2}gx_{out}Fx_{in}
-\frac{\imath}{2}
e_{0}{\mathcal M}^2-\frac{\imath}{2}\frac{s^2}{e_{0}}\right.
$$
$$
\left.
-\frac{ \imath }{4}(x_{out}-x_{in})gF
\coth\frac{ge_{0}F}{2}(x_{out}-x_{in})\right\}.
\eqno(13)
$$

If the electromagnetic field is turned off we obtain again 
the result Eq. (8) with
 $s=0$ and, for $\alpha=0$, we obtain the path integral
  representation for 
 the spinless particle propagator in $(2+1)$-dimensions.
 
 In summary, we have obtained a path integral representation
  for the propagator of a free 
 anyon Eq. (8) and its representation in momentum space 
 Eq. (9) taking into account a 
 continuous family of Lagrangians Eq. (4). These ones were 
 obtained by a
  group-theoretical 
 approach with minimal extension that preserves all canonical structure 
 of the space-time and spin algebra. We also consider an anyon in a 
 constant electromagnetic field. To obtain the propagator, 
 a convenient gauge $\;$ ${\dot{\sigma}}=0$,$\;$${\dot e}=0
  $ was considered.

\end{document}